\documentclass{aa}  
\usepackage{graphicx}
\usepackage{color}
\usepackage{url}
\usepackage{hyperref}
\usepackage[normalem]{ulem}
\usepackage{xspace}
\usepackage{txfonts}
\usepackage{natbib}
\newcommand{\psrtar}{IGR\,J17498$-$2921}

\newcommand{\gtap}{\mathrel{\hbox{\rlap{\lower.55ex \hbox {$\sim$}}
                   \kern-.3em \raise.4ex \hbox{$>$}}}}
\newcommand{\ltap}{\mathrel{\hbox{\rlap{\lower.55ex \hbox {$\sim$}}
                   \kern-.3em \raise.4ex \hbox{$<$}}}}

\newcommand{\nustar}{{NuSTAR}\xspace}
\newcommand{\nicer}{{NICER}\xspace} 
\newcommand{\cxo}{{\it Chandra}\xspace}

\newcommand{\Integ}{{INTEGRAL}\xspace} 
\newcommand{\maxi}{{MAXI}\xspace} 

\newcommand{\rxte}{{\it RXTE}\xspace}
\newcommand{\hxmt}{{Insight-HXMT}\xspace}

\def\be{\begin{equation}} 
\def\ee{\end{equation}}

\begin{document}
   \title{
   Broad-band X-ray spectral and timing properties of the accreting millisecond X-ray pulsar IGR J17498--2921 during the 2023 outburst}
   \titlerunning{High-energy characteristics of IGR J17498--2921}
   \authorrunning{Z. Li et al.}


   \author{Z.S. Li\inst{1}
           \and
              L. Kuiper\inst{2}
            \and
              Y.Y. Pan\inst{1}
     \and
      M. Falanga\inst{3,4}
             \and
    J. Poutanen\inst{5,6}
    \and
        Y.P. Chen\inst{9}
    \and
    R.X. Xu\inst{7,8}
    \and
    M.Y. Ge\inst{9}
    \and
    Y. Huang\inst{9}
    \and
    L.M. Song\inst{9}
    \and
    S. Zhang\inst{9}
    \and
    F.J. Lu\inst{9}
    \and
    S.N. Zhang\inst{9}
          }
   \offprints{Z. Li}

   \institute{Key Laboratory of Stars and Interstellar Medium, Xiangtan University, Xiangtan 411105, Hunan, P.R. China\\
              \email{lizhaosheng@xtu.edu.cn}
               \and
            SRON-Netherlands Institute for Space Research, Sorbonnelaan 2,  3584 CA, Utrecht, The Netherlands
              \and
              International Space Science Institute (ISSI), Hallerstrasse 6, 3012 Bern, Switzerland
              \and
              Physikalisches Institut, University of Bern, Sidlerstrasse 5, 3012 Bern, Switzerland
              \and
              Department of Physics and Astronomy, FI-20014 University of Turku, Finland
              \and
              Space Research Institute of the Russian Academy of Sciences, Profsoyuznaya str. 84/32, 117997 Moscow, Russia
              \and
              Department of Astronomy, School of Physics, Peking University, Beijing 100871, People's Republic of China
              \and
              Kavli Institute for Astronomy and Astrophysics, Peking University, Beijing 100871, People's Republic of China
              \and
              Key Laboratory of Particle Astrophysics, Institute of High Energy Physics, Chinese Academy of Sciences, 19B Yuquan Road, Beijing 100049, China
              }

   \date{Received xx  / Accepted xx}

  \abstract{
We report on the broadband spectral and timing properties of the accreting millisecond X-ray pulsar \psrtar\ during its April 2023 outburst using data from \nicer\ (1--10 keV), \nustar\ (3--79 keV), \hxmt\ (2--150 keV), and \Integ\ (30--150 keV). We detect significant 401 Hz pulsations across the 0.5--150 keV band. The pulse fraction increases from $\sim$2\% at 1~keV to $\sim$13\% at 66~keV.  Five type-I X-ray bursts have been detected, including three photospheric radius expansion bursts, with a rise time of $\sim$2~s and an exponential decay time of $\sim$5~s. The recurrence time is $\sim$9.1~h, which can be explained by unstable thermonuclear burning of hydrogen-deficient material on the neutron star surface. The quasi-simultaneous 1--150~keV broadband spectra from \nicer, \nustar\ and \Integ\ can be well fitted by an absorbed reflection model, \texttt{relxillCp}, and a Gaussian line of instrumental origin. The Comptonized emission from the hot corona is characterized by a photon index $\Gamma$ of $\sim$1.8 and an electron temperature $kT_{\rm e}$ of $\sim$40~keV. We obtain a low inclination angle $i\sim34\degr$. The accretion disk shows properties of strong ionization, $\log(\xi/{\rm erg~cm~s^{-1}})\sim4.5$, over-solar abundance, $A_{\rm Fe}\sim  7.7$, and high density, $\log(n_{\rm e}/{\rm cm^{-3}})\sim 19.5$. However, a lower disk density with normal abundance and ionization could also be possible. From the inner disk radius $R_{\rm in}=1.67R_{\rm ISCO}$ and the long-term spin-down rate of $-3.1(2)\times10^{-15}~{\rm Hz~s^{-1}}$, we constrain the magnetic field of \psrtar\ in the range of $(0.9-2.4)\times10^8$ G.

}

   \maketitle

\section{Introduction}
\label{sec:intro}

Accreting millisecond X-ray pulsars (AMXPs) are binary systems hosting a rapidly  rotating, old ($\sim$Gyr) neutron star (NS) possessing a relatively weak magnetic field of $10^{8-9}$~G  and having a low-mass companion. AMXPs are identified based on their coherent X-ray pulsations at spin frequencies higher than $\sim100$ Hz during X-ray outbursts \citep[see][for reviews]{Campana2018,DiSalvo2020,papitto20,Patruno2021}. On average, about one new AMXP is discovered per year since the first confirmed source SAX J1808.4--3658. 
Among the recent ones, we can mention, MAXI J1816--195 \citep{Bult22,Li23} and MAXI J1957+032 \citep{Sanna22} in 2022 and SRGA J144459.2--604207 \citep{2024ATel16480,2024ATel16548,Ng24,Molkov24} in 2024. Usually, a few outbursts from the known AMXP sample can be observed each year too. 

\psrtar\ has been discovered by \Integ\ during its 2011 outburst, and confirmed as an AMXP with the spin frequency of $\sim$401~Hz, orbital period of 3.8~hr, and projected semi-major axis, $a_{\rm X}\sin i/c\approx0.365$ lt-s \citep{Papitto11}. The source position using \cxo\ X-ray observation was determined at $\alpha_{\rm 2000} = 17^{\rm h}49^{\rm m}55\fs35$ and $\delta_{\rm 2000} = -29\degr19\arcmin19\farcs6$ with an uncertainty of $0\farcs6$ at 90\% confidence level \citep{2011ATel.3606....1C}.  The coherent pulsations have been detected up to 65~keV with an energy-independent pulsed fraction of 6--7\%. The pulse profiles are well described by a sine wave  and the soft lags of $\sim-60~\mu$s have been detected which saturated near 10~keV \citep{falanga12}.

During the 2011 outburst of \psrtar, type-I X-ray bursts have been discovered, including one photospheric radius expansion (PRE) burst observed by \rxte\ \citep{2011ATel.3568....1L,falanga12}. The distance to the source was estimated at 8 kpc. From the burst profiles and recurrence times one could conclude that they were powered by the unstable burning of pure helium or the material with the CNO metallicity up to twice solar abundance. The burst oscillation signal around its spin frequency has been detected mainly during the cooling tail \citep{Chakraborty12}.

In this work, we study the \nicer, \nustar, \hxmt, and \Integ\ observations of \psrtar\ during its 2023 outburst. The data analysis and the outburst profile are introduced in Sects.~\ref{sec:obs} and \ref{sec:outburst}, respectively. The spectral evolution during \nicer\ observations and the broadband spectra are reported in Sect.~\ref{sec:spec}. We investigate the timing properties of \psrtar\  in Sect.~\ref{sec:timing}. 
The X-ray burst properties are presented in Sect.~\ref{sec:bursts}. 
We discuss all the results in Sect.~\ref{sec:disc}


\begin{figure*} 
\centering
\includegraphics[width=\textwidth]{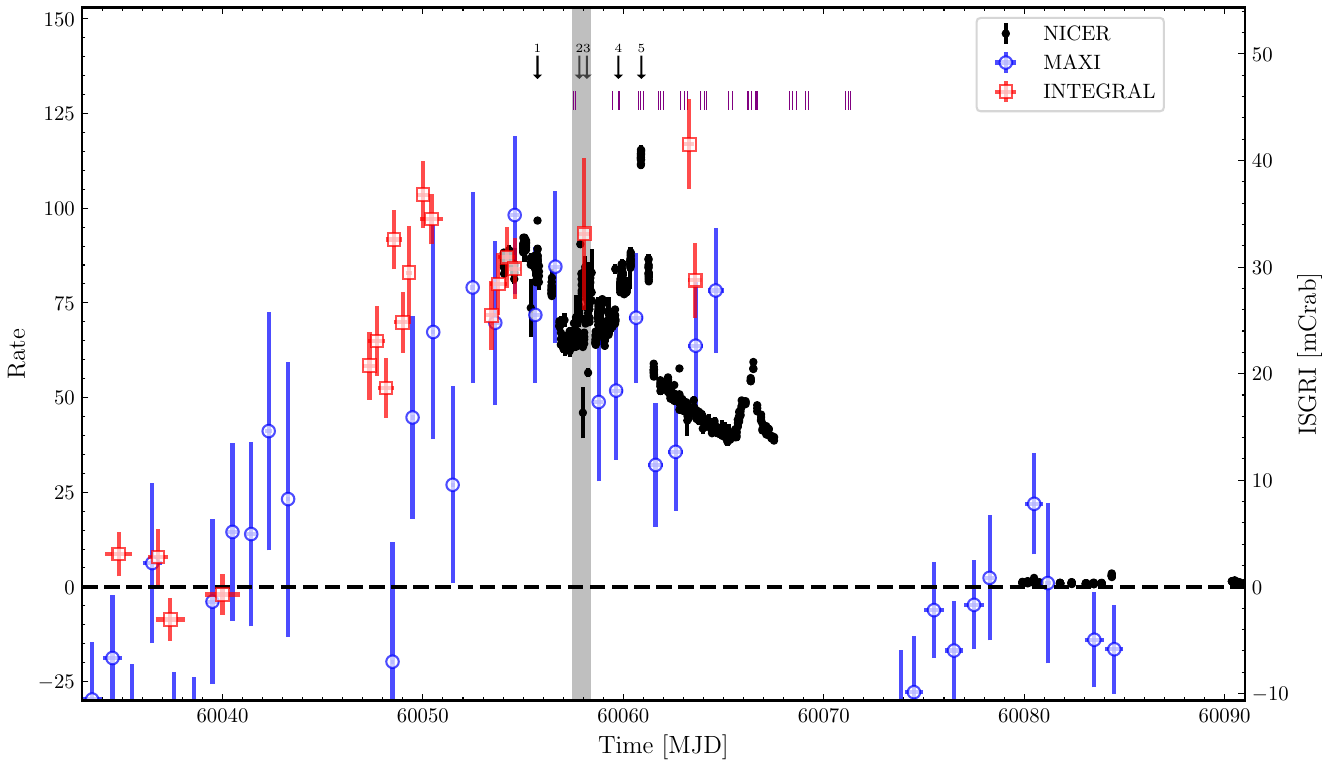}

\caption{Light curves of \psrtar\ from \nicer\ (black dot, 0.5--10 keV, in units of cnt~$\rm{s^{-1}}$), \maxi/GSC (blue open circle, 2.0--6.0 keV), and \Integ/ISGRI (red open square, 20-60 keV) during its 2023 outburst.  
The \maxi\ rate is in units of  ph~cm$^{-2}$~s$^{-1}$ multiplied by 560 to overlap with the \nicer\ data. The scale of the 20--60 keV ISGRI light curve is shown along the right vertical axis and is given in units of mCrab. The grey area presents the epoch of \nustar\ observations. The black arrows on the upper center mark the start time of five type I X-ray bursts with burst number. 
The horizontal dashed line shows the background level. The vertical purple lines present the start time of \hxmt\ observations. The count rates of the three \hxmt\ instruments were polluted by several nearby (Galactic center) sources, and therefore, not included here.  }
\label{fig:outburst}
\end{figure*}


\section{Observations}
\label{sec:obs}

After being twelve years in quiescence, \psrtar\ showed new activity cycle as was first noticed by \citet{2023ATel15996} using \Integ\ data on 2023 April 13--15. \nicer, \hxmt, and \nustar\ had carried out dedicated follow-up observations, covering the main epoch of the outburst \citep{2023ATel15998}. We collected data from \psrtar\ observed by these instruments as well as \Integ\ during its April 2023 outburst. 

\subsection{\Integ}
\label{sec:integral}

In this work, we used mainly data from the coded mask soft gamma-ray imager IBIS/ISGRI \citep{ubertini03,lebrun03} aboard \Integ\ \citep{winkler03} at energies from 20 to 150 keV. For ISGRI, the data were extracted for all pointings with a source position offset $\leq14\fdg5$. The data reduction was carried out using the standard Offline Science Analysis (OSA) version 11 distributed by the INTEGRAL Science Data Center \citep{courvoisier03}. 

\Integ\ Galactic centre ToO observations --  targeting  Sgr~A* and GX 5$-$1 -- during satellite revolution (Rev.) 2628 on 2023 April 13--15 (MJD 60047.181--60049.378) showed renewed activity of AMXP \psrtar\ \citep{2023ATel15996} in IBIS-ISGRI data. 
A zoom-in on Rev.~2628 revealed that \psrtar\ reached already a $8.3\sigma$ detection in the 20--60 keV band during the 25.2 ks ToO observation of Sgr~A* performed at the beginning of Rev.~2628 on MJD 60047.181--60047.441.
Since the onset of the outburst \psrtar\ was in the \Integ\ field-of-view during Galactic centre observations performed throughout revolutions 2628 (176.4 ks; $28.4\sigma$), 2629 (73.8 ks; $27.8\sigma$), 2630 (119.4 ks; $29.1\sigma$), 2632 (50 ks; $10.5\sigma$), and 2634 (100 ks; $16.2\sigma$) covering the period 2023 April 13--30,  (MJD 60047.181--60064.39). Significances were quoted for the 20--60 keV band. We also analysed the pre-outburst 2023 Galactic centre observations performed during  Revs. 2623--2625 (311.2 ks), clearly indicating that \psrtar\ was off. 

For the spectral analysis, we extracted a (outburst averaged) combined spectrum from the mosaic of Revs. 2628--2630/2632/2634 observations during which the source showed a relatively flat outburst profile with the source flux being about 30 mCrab (see Fig.~\ref{fig:outburst}).

\subsection{NICER}
\label{sec:nicer}

\nicer\ provides non-imaging detectors with a field of view (FOV) of 30 arcmin$^2$ and an absolute timing accuracy of  100~ns.  About one week after the detection of the 2023 outburst of \psrtar\ by \Integ,   \nicer\ carried out follow-up observations starting on 2023 April 20 (MJD 60054). The last observation ended on 2023 July 8 (MJD 60134), when the source was returned to a quiescent state. In total, 46 \nicer\ observations are available, including Obs IDs 6203770101$\sim$6203770105 and 6560010101$\sim$6560010141. The total exposure time is 130~ks, including 81~ks in the outburst, i.e., the first 18 observations from April~20 to May~3, and 49~ks in quiescence, i.e., the last 28 observations from May~15 to July~8. 

The \nicer\ data were processed by using \textsc{heasoft} V6.31 and the \nicer\ Data Analysis Software (NICERDAS) version~10. The standard filtering criteria were applied, including an angular offset of the source \texttt{ANG\_DIST} less than $0\fdg015$, an Earth limb elevation angle and bright Earth limb angle larger than 20\degr\ and 30\degr, respectively, an undershoot rate of \texttt{underonly\_range}=0--500, an overshoot rate of \texttt{overonly\_range}=0--30 and a \nicer\ location outside the South Atlantic Anomaly. We extracted 1~s binned light curves in the 0.5--10 and 12--15~keV bands using the command \texttt{nicerl3-lc}. Time intervals showing flaring background in 12--15~keV light curves were discarded from further analysis.  

We identified one type I X-ray burst in Obs. Id. 6560010101 (see Sect.~\ref{sec:bursts}). The spectra were generated using \texttt{nicerl3-spect} tool, and the associated ancillary response files (\texttt{ARFs}), response matrix files (\texttt{RMFs}), and 3C50 background spectra \citep{Remillard22} were produced simultaneously.

In the timing analysis we used a multi-mission serving barycentering tool adopting the JPL DE405 Solar System ephemeris, written in \texttt{IDL} and developed at SRON, which is equivalent and compatible with \textsc{heasoft} tool \texttt{barycorr}.
To estimate background corrected pulsed emission properties (see Sect.~\ref{sec:pulse_fraction_lag}), we used \nicer\ data collected during Obs. Ids. 6560010114--6560010119 (2023 May 15--20; MJD 60079.9--60084.4; 9.389~ks GTI time)\footnote{Observations performed before the `optical light leak' period that commenced on May 22, 2023 between 13:00--14:00 UTC.} as background reference sample when the source was in a well-established Off state (see Fig. \ref{fig:outburst}).

\subsection{NuSTAR}
\label{sec:nustar}

\nustar\ has observed \psrtar\ on  2023 April 23 for a total exposure time of  44.6~ks (Obs. ID 90901317002, MJD 60057.44--60058.48). The event files from FPMA and FPMB have been cleaned by using the \nustar\ pipeline tool \texttt{nupipeline}.

The source light curves were extracted from a circle region with a radius of $100\arcsec$ centered on the source location using \texttt{nuproducts}. The count rate from \nustar\ maintained a constant level of $\sim$19~cnt~s$^{-1}$ in the 3--79 keV band, except for two type I X-ray bursts that showed up during the \nustar\ observation. 
We excluded these two X-ray bursts in producing the source spectra, response, and ancillary response files to perform joint spectral fitting with \nicer\ spectra (see Sect.~\ref{sec:nicer_nustar}). The background spectra were obtained from a source free circular region with a radius of $100\arcsec$ centered at $(\alpha_{2000},\delta_{2000})= (17^{\rm h}50^{\rm m}20\fs94, -29\degr19\arcmin18\farcs95)$. 
We also generated time-resolved burst spectra for the two X-ray bursts with the option \texttt{usrgtifile}  applied in the command \texttt{nuproducts}. 

In timing studies we used barycentered events adopting the JPL DE405 Solar System ephemeris and applying fine-clock-correction file \#169, from a circular region with a radius of $90\arcsec$ centered on the \cxo\ X-ray location \citep{2011ATel.3606....1C} of \psrtar\ as source sample. To obtain background corrected quantities like pulsed fraction (see Sect. \ref{sec:pulse_fraction_lag}) we used a source free circular region (located on the same chip as our target) with a radius of $90\arcsec$ centered at $(\alpha_{2000},\delta_{2000})= (17^{\rm h}49^{\rm m}53\fs412, -29\degr23\arcmin19\farcs522)$ as a background reference sample.

\subsection{\hxmt}
\label{sec:hxmt}

The \hxmt\ \citep[Insight Hard X-ray Modulation Telescope,][]{hxmt} has three slat-collimated and non-imaging telescopes, the Low Energy X-ray telescope \citep[LE, 1--12 keV; ][]{hxmt-le}, the Medium Energy X-ray telescope \citep[ME, 5--35 keV; ][]{hxmt-me}, and the High Energy X-ray telescope \citep[HE, 20--350 keV; ][]{hxmt-he}, which have the capabilities of broadband X-ray timing and spectroscopy. The FOV of \hxmt\ is approximately $6^\circ\times6^\circ$. The absolute timing accuracy of \hxmt\ has been verified from the alignment of the pulse profiles in the AMXP MAXI J1816--195, i.e., within 14~$\mu$s between \hxmt/ME and \nicer, and $0.9\pm13.9$~$\mu$s between \hxmt\ ME and HE, in the same energy bands  \citep{Li23}. \hxmt\ has carried out twelve ToO observations (PI, Z. Li, Obs. ID P0504093001--P0504093012) between April 23 and  May 7,  2023. We analyzed the data using the \hxmt\ Data Analysis Software (HXMTDAS V2.05). The LE, ME, and HE data were calibrated by using the scripts \texttt{lepical}, \texttt{mepical}, and \texttt{hepical}, respectively.  For each instrument, the good time intervals (GTIs) were individually obtained from the scripts \texttt{legtigen}, \texttt{megtigen}, and \texttt{hegtigen}, resulting in the exposure time of 13.6, 89.6, and 70.3~ks from LE, ME, and HE, respectively.  

We obtained background-subtracted light curves from LE, ME and HE in 2--10, 10--35 and 27--250~keV, respectively. Two type I X-ray bursts were observed, the first from Obs ID P050409300302 with LE and ME data available, while the second from Obs ID P050409300402 only from ME data. After ignoring these X-ray bursts, the count rates remained at constant levels of 51.6, 15.7, and 33.5~cnt~s$^{-1}$ for LE, ME and HE, respectively. However, the \nicer\ light curve showed a reduction in count-rate of at least a factor of two during the \hxmt\ observations. Therefore, due to its large FOV the \hxmt\ LE, ME and HE light curves were strongly affected by the nearby bright sources, e.g. the persistent black hole candidate 1E 1740.7--2942 located at $1\fdg 37$ from the source position. Therefore, we did not perform a joint spectral fitting using the \hxmt\ LE/ME/HE data (see Sect.~\ref{sec:spec}). The nearby bright sources did not exhibit pulsation signatures around 401~Hz, and their emissions only contributed to a higher (statistically flat) background level in the pulsation studies.

We used the timing data from \hxmt\ ME and HE, barycentered using the tool \texttt{hxbary} with the JPL DE405 Solar System ephemeris, to study the X-ray pulsation of \psrtar. For the X-ray bursts, we extract the pre-burst spectra and the time-resolved burst spectra via the tools \texttt{lespecgen} and \texttt{mespecgen}, and their response matrix files are generated from \texttt{lerspgen} and \texttt{merspgen},  for LE and ME, respectively. See Sect.~\ref{sec:bursts} for the details of the X-ray burst studies.



\section{Outburst profile}
\label{sec:outburst}

In Fig.~\ref{fig:outburst}, we show the outburst profile of \psrtar\ during the 2023 outburst, which includes the observations from \nicer\ (0.5--10 keV), \maxi\ (2.0--6.0 keV), \Integ/ISGRI (20--60 keV). The \hxmt\ light curves are not presented in Fig.~\ref{fig:outburst} due to the pollution of nearby sources, but the start time of \hxmt\ observations are marked as vertical purple lines. The observations of the outburst rise were poorly covered.  By interpolating the \Integ/ISGRI light curve, the onset of the outburst was determined around MJD 60040. At the same time, the \maxi\ rate was still consistent with the background, indicating a hard spectral state at the beginning of the outburst. The source reached its peak within 5 days. In the next $\sim10$ days, from the \nicer\ 0.5--10 keV light curves, \psrtar\  had experienced complicated fluctuations, which exhibited three main peaks on MJD 60055.1, 60058.4, and 60060.8 with the rate of $\sim$90, $\sim$85, and $\sim$110~cnt~s$^{-1}$, respectively. Starting after the third peak, the count rate decreased slowly to quiescent. During this process, a reflare in the later stage of the outburst appeared on MJD 60066.5 with a peak rate of $\sim$60~cnt~s$^{-1}$. From the \maxi\ data, the source returned to the quiescent state before MJD 60074, i.e., the outburst lasted around one month. During the quiescent state, the \nicer\ count rate was consistent with the background level without distinct reflares.  The onset time of five type I X-ray bursts are labeled as vertical arrows in Fig.~\ref{fig:outburst}. 



\section{Spectral analysis}
\label{sec:spec}

We used \textsc{xspec} version 12.13.0c \citep{arnaud96} to fit the persistent spectra from \psrtar. The Tübingen–Boulder absorption model, \texttt{tbabs}, with abundances from \citet{wilms00}, was applied to account for the interstellar medium absorption.  There are several quasi-simultaneous observations between \nicer\ and \hxmt. The joint \nicer\ and \hxmt\ spectral fitting revealed that the \hxmt\ fluxes were 3--5 times higher than the \nicer\ flux. Similar to the count rate from \hxmt, the spectra reconfirm that a nearby source contributed more than \psrtar\ in the \hxmt\ data. Therefore, we only fitted \nicer\ spectra for the whole outburst in Sect.~\ref{sec:nicer_spec}, and performed the joint \nicer, \nustar, and \Integ\ spectral fitting in Sect.~\ref{sec:nicer_nustar}. The uncertainties of all parameters are quoted at the $1\sigma$ confidence level.

\begin{figure} 
\centering
\includegraphics[width=0.5\textwidth]{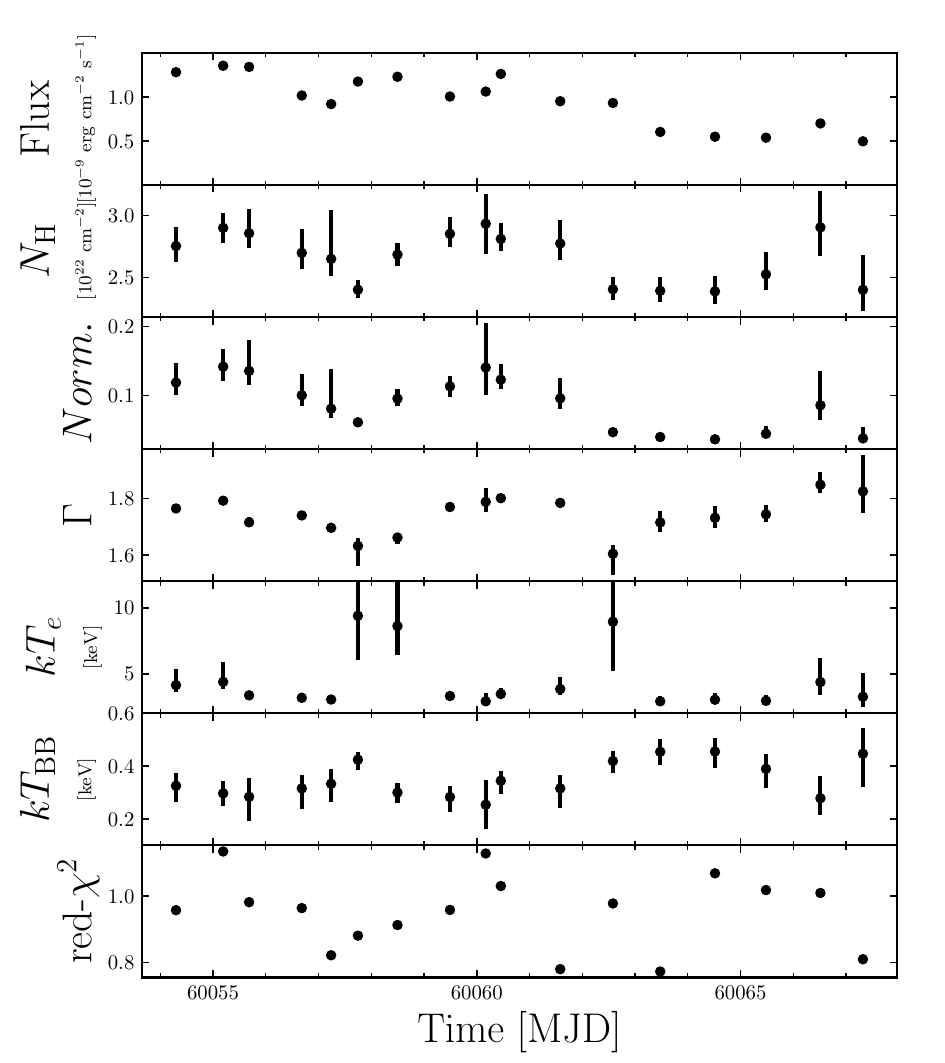}
\caption{The best-fitted parameters of \nicer\ spectra by using the model \texttt{tbabs*(Gaussian+nthcomp)}. }
\label{fig:nicer_spectra}
\end{figure}

\subsection{\nicer\ spectra}
\label{sec:nicer_spec}


We fitted the \nicer\ spectra in between MJD 60054.3--60067.3. After MJD 60067.3, no \nicer\ spectra were available until the source returned to the quiescent state. Below 1 keV, the \nicer\ spectra showed substantial excess possibly due to instrumental origin, therefore, we focused on the 1--10 keV energy range. We fitted the continuum with a thermal Comptonized component, \texttt{nthcomp}, modified by photoelectric absorption modeled by \texttt{tbabs}. The residuals also showed an emission feature at 1.6--1.7 keV from the instrument. We added a Gaussian component to account for it. The whole model is \texttt{tbabs$\times$(Gaussian+nthcomp)}. The parameters include the hydrogen column density, $N_{\rm H}$, for \texttt{Tbabs}, the line energy, width, and normalization for \texttt{Gaussian}, the asymptotic power-law photon index, $\Gamma$, the electron temperature, $kT_{\rm e}$, the seed photon temperature, $kT_{\rm bb}$, the type of the seed photons and the normalization for \texttt{nthcomp}. We set the type for the seed photons to a blackbody distribution. The bolometric flux was estimated via \texttt{cflux} in 1--250 keV. The uncertainties were obtained from the command \texttt{error}.

The best-fitted parameters are shown in Fig.~\ref{fig:nicer_spectra}. The bolometric flux showed fluctuations between $(0.92-1.35)\times10^{-9}~{\rm erg~s^{-1}~cm^{-2}}$ for the first $\sim8$ days, and then slowly decayed to $\sim5\times10^{-10}~{\rm erg~s^{-1}~cm^{-2}}$  within one reflare in next $\sim7$ days. 
We found increasing and then decreasing variations of the hydrogen column density, $N_{\rm H}\sim(2.4-2.9)\times10^{22}~{\rm cm^{-2}}$, and the power-law photon index, $\Gamma\sim1.6-1.8$, and opposite trends of the electron temperature, $kT_{\rm e}\sim2.9-9.4$ keV, and the seed photon temperature, $kT_{\rm BB}\sim0.25-0.46$ keV, with a quasi-periodical timescale of $4\sim5$ days. 

\begin{figure} 
\centering
\includegraphics[width=0.46\textwidth]{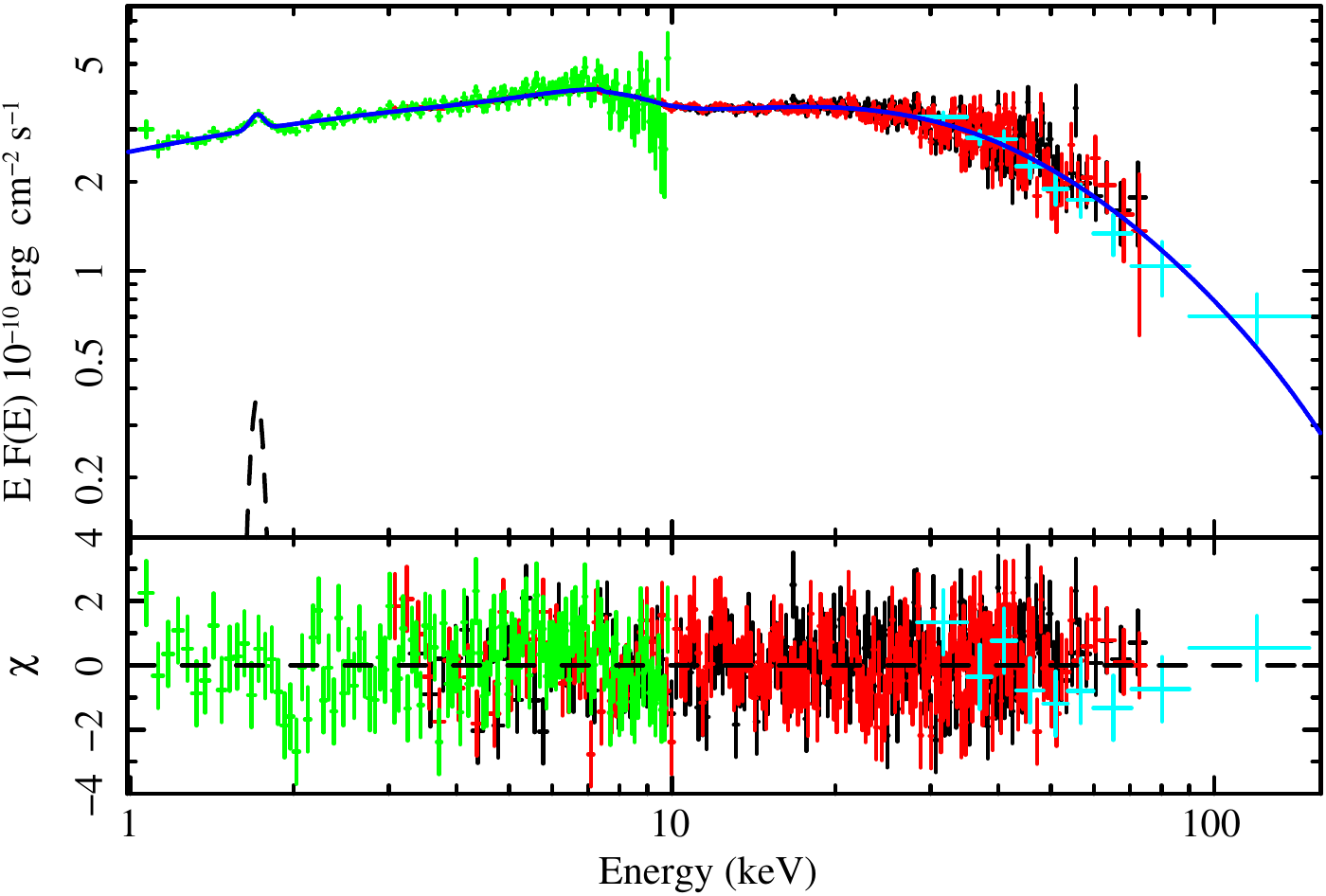}
\caption{The joint spectral fitting of \nicer, \nustar\ and \Integ/IBIS-ISGRI. The spectra were observed during MJD 60057.51--60057.53 for \nicer\ (green), MJD 60057.44--60058.48 for \nustar\ (red for FPMA and black for FPMB), and  MJD 60047.4--60064.4 for  \Integ/IBIS (cyan). The black dashed line shows the Gaussian component at 1.7 keV, likely an instrumental feature. The blue solid line represents the best-fitted model \texttt{tbabs$\times$(Gaussian+relxillcp)}. }
\label{fig:broadband_spec}
\end{figure}

\subsection{Broadband spectral fitting}
\label{sec:nicer_nustar}

We performed a quasi-simultaneously joint spectral fitting between \nicer, \nustar\ and \Integ\ for \psrtar, covering the energy range of 1--150 keV. The spectra were collected between MJD 60057.51--60057.53 for NICER Obs. ID 6203770103, MJD 60057.44--60058.48 for \nustar. We also produced the average INTEGRAL/IBIS-ISGRI spectrum during MJD 60047.16--60064.39 to guarantee a high signal-to-noise ratio.   We introduced a multiplication factor for each instrument to account for the cross-calibration uncertainties and possible flux variations. The factor was fixed at unity for \nustar/FPMA and set free for other instruments.

We first attempted to fit the joint spectra using the model as mentioned above \texttt{tbabs*(nthcomp+gaussian)}. The model can not explain the broadband spectra well with $\chi^2_{\nu}\approx4.0$ for 565 degrees of freedom (dof). We also tried the thermal Comptonization model, \texttt{compps} \citep{ps96}, in the slab geometry, which has been applied to its 2011 outburst \citep{falanga12}, but the fit was also very poor, $\chi^2_{\nu}\approx3.0$ for 563 dof \citep[see also][]{falanga05,falanga05b,falanga08,falanga11,falanga12,deFalcoa,deFalcob,ZLI2018,Kuiper20,ZLi21,Li23}. We note that the residuals show a reflection feature. Therefore, we replaced \texttt{nthcomp} with the relativistic reflection model \texttt{relxillCp}, where the incident spectrum is modeled by an \texttt{nthcomp}  Comptonization continuum \citep[see ][for more details]{Ludlam24}. The free parameters of the model are as follows: the inclination of the system, $i$, the inner radius of the disc, $R_{\rm in}$, in units of the inner-most stable circular orbit (ISCO, $R_{\rm ISCO}$), the power law index of the incident spectrum, $\Gamma$, the electron temperature in the corona, $kT_{\rm e}$, the logarithm of the disk ionization, $\log\xi$, the iron abundance normalized to the Sun, $A_{\rm Fe}$, the density of the disk in logarithmic units, $\log n_{\rm e}$, and the reflection fraction, $f_{\rm refl.}$. We fixed the
inner and outer emissivity indices, $q_1$ and $q_2$, both at 3, the break radius
between these two emissivity indices and the outer disk radius, $R_{\rm out}=R_{\rm break}=1000R_{\rm g}$, where $R_{\rm g}=GM_{\rm NS}/c^2$ is the gravitational
radius, $G$ and $c$ are the gravitational constant and the speed of light, respectively. Assuming the NS mass and radius of $1.4M_\odot$ and 10 km, respectively, for \psrtar\ spinning at 401 Hz, we obtained the dimensionless spin parameter $a = 0.188$, which was also fixed. The model fitted the spectra acceptably with $\chi^2_{\nu}\approx1.12$ for 560 dof.  We tried to add an extra thermal component, \texttt{bbodyrad} from the NS surface, or \texttt{diskbb} from the accretion disk, to the model, however, the fit did not improve significantly, i.e., $\Delta\chi^2<3$ with two more free parameters.  Therefore, we propose that no significant thermal emissions from the disc or NS surface are found in the spectra.  Moreover, no apparent features appeared in the residuals (see Fig.~\ref{fig:broadband_spec}). 

\begin{figure*}[t]
\centering
\includegraphics[width=0.9\textwidth]{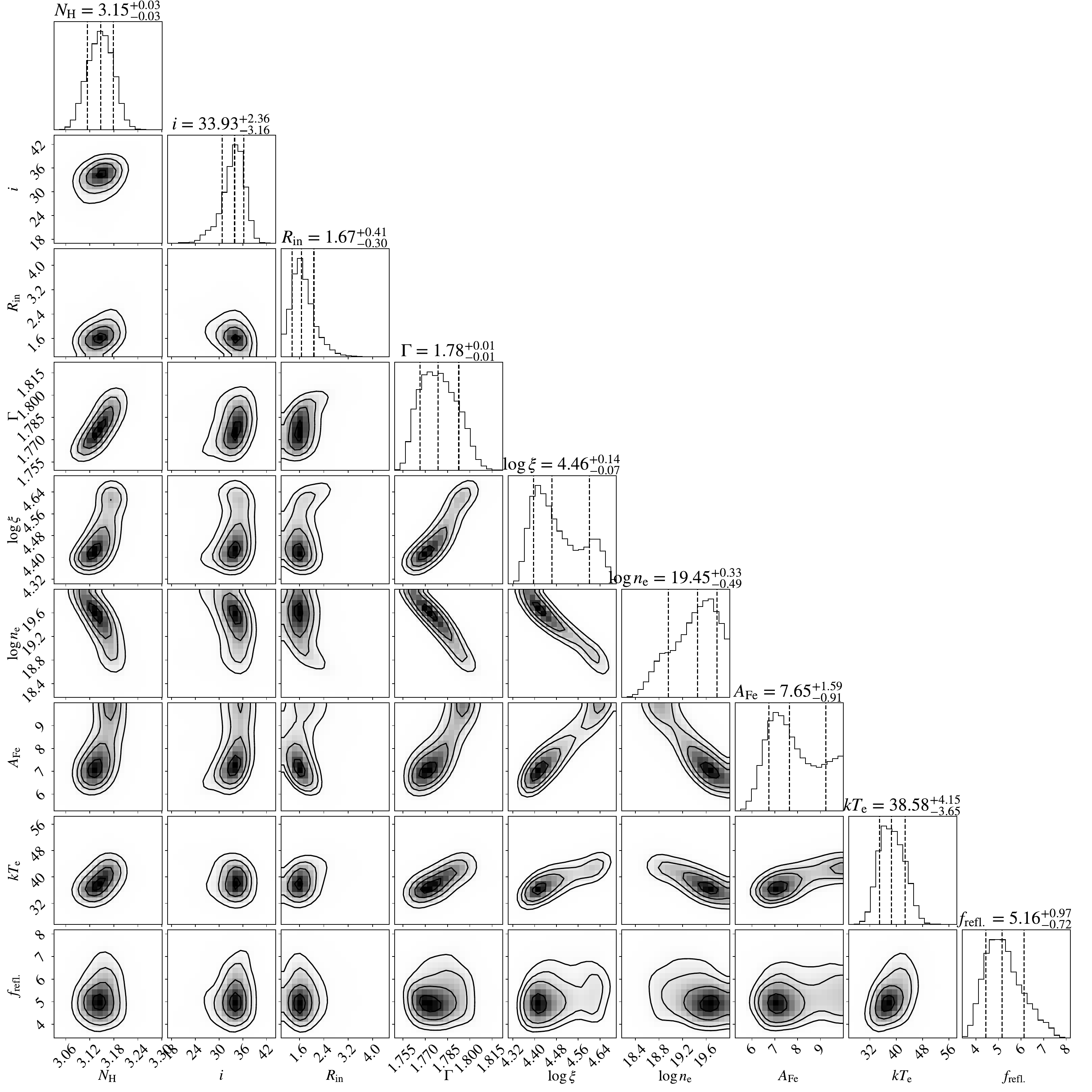}

\caption{ The posterior distributions of the fitted parameters obtained by MCMC simulations. The parameters of the Gaussian component and the normalization of \texttt{relxillCp} are not shown here.}
\label{fig:mcmc}
\end{figure*}

We applied the Goodman–Weare algorithm of Monte Carlo Markov Chain (MCMC) to investigate the best-fit parameters further to produce contour plots \citep{Goodman10}. We adopted 100 walkers and a chain length of $5\times10^6$ to calculate the marginal posterior distributions. The first $10^5$ steps were set as burn-in length and discarded. The contours are presented in Fig.~\ref{fig:mcmc}, which are produced by using the procedure \texttt{corner.py} \citep{Foreman-Mackey16}. The $1\sigma$ confidence levels of the best-fit parameters from MCMC are listed in Table~\ref{table:spec}. The unabsorbed bolometric flux is $(1.45\pm0.01)\times10^{-9} ~{\rm erg ~s^{-1}~ cm^{-2}}$ in 1--250 keV, which is larger than the result solely from the \nicer\ spectrum reported in Sect.~\ref{sec:nicer_spec} because of using different datasets and the models. The $N_{\rm H}$ is $(3.15\pm0.03)\times10^{22}~{\rm cm^{-2}}$, slightly higher than the results from  the \nicer\ spectra, but consistent with the value reported by \citet{2011ATel.3555....1B}. All multiplication factors are close to unity, suggesting proper calibration between all instruments and no strong variability during the quasi-simultaneous observations. The obtained values characterize the accretion disk of strong ionization, over-solar abundance, and high density. The Comptonized emission associated with a hot corona is characterized by the photon index of $  1.78\pm0.01$ and the electron temperature of $kT_{\rm e}=   39\pm 4$ keV, implying a hard spectral state. The inclination angle of the accretion disk is $34\pm4$ deg, in agreement with the absence of dips or eclipses in the light curves. From the relation $R_{\rm ISCO}=6GM_{\rm NS}/c^2[1-a(2/3)^{3/2}]$ \citep{Miller98}, the inner disk radius $R_{\rm in}=1.67R_{\rm ISCO}$ corresponds to $18.6_{-3.3}^{+4.5}$ km, suggesting that the accretion disk is located rather closed to the NS surface for a typical radius of 10 km. On the other hand, the obtained inner disk radius set an upper limit of the NS radius \citep[see][and references therein]{Ludlam24}, which is consistent with most NS equations of state \citep[see][and references therein]{Burgio21}. 

 \begin{table} 
 \caption{\label{table:spec} Best-fit spectral parameters  of    the \nicer/\nustar/\Integ\ data for \psrtar\ for the model  {\tt constant$\times$tbabs$\times$(gaussian+relxillCp)}.}
 \centering
 \begin{tabular}{ll} 
  \hline 
 \hline 

Parameter (units) &Best-fit values  \\

 \hline 
$N_{\rm H}~(10^{22}~ {\rm cm}^{-2})$ & $   3.15\pm0.03$ \\
$E_{\rm line}$   (keV)              & $   1.71\pm0.02$ \\
$\sigma$   (keV)                    & $ 0.054\pm0.026$ \\
Norm($\times10^{-3}$)               & $ 0.97_  {-0.25}^{+0.30   }$ \\
$i$ (deg)                           & $   34_{-3}^{+2}$ \\
$R_{\rm in} (R_{\rm ISCO})$         & $  1.7_{-0.3}^{+ 0.4}$ \\
$\Gamma$                            & $   1.78\pm0.01$ \\
$\log(\xi/{\rm erg~cm~s^{-1}})$     & $   4.46_{-0.07}^{+0.14}$ \\
$\log(n_{\rm e}/{\rm cm^{-3}})$           & $   19.5_  {-0.5}^ {+0.3}$ \\
$A_{\rm Fe}$                        & $   7.7_{-0.9}^ {+1.6   }$ \\
$kT_{\rm e}$      (keV)             & $   39\pm4$ \\
$f_{\rm refl.}$                     & $   5.2_  {-0.7}^ {+1.0}$ \\
Norm$_{\rm refl.}~(\times10^{-4})$  & $2.8\pm 0.4$ \\
$C_{\rm \nustar/FPMA}$              & 1 (fixed) \\ 
$C_{\rm \nustar/FPMB}$              & $   1.01\pm0.01 $ \\
$C_{\rm \nicer}$                    & $  0.97\pm0.01 $ \\
$C_{\rm \Integ/ISGRI}$               & $   1.19\pm0.03 $ \\
  \hline 
 $\chi^{2}/{\rm d.o.f.}$ &  625.49/560 \\ 
 $F_{\rm bol}$ ($10^{-9}$ erg s$^{-1}$ cm$^{-2}$)$^{(a)}$ & $1.45\pm0.01$  \\ 
 \hline  
 \end{tabular}  



\tablefoot{
 The multiplication factor for all instruments is provided.
\tablefoottext{a} {Unabsorbed flux in the 1--250 keV energy range.}\\
}
 \label{tab:broadband_spec} 
 \end{table} 

We explored the possibility of lower solar abundance by fixing $A_{\rm Fe}$ at two times solar abundance. However, the model fitted the spectra poorly with $\chi^2_\nu=1.59$ for 561 dof. Adding an extra blackbody component with a temperature of $1.6\pm0.1$ keV and normalization of $0.7_{-0.2}^{+0.3}~{\rm km^2}$, the best-fitted results improved to $\chi^2_\nu=1.40$ for 559 dof, which was also worse than the model described above. We noticed that only the inner disk radius, $R_{\rm in}=4.5_{-2.1}^{+10.7}R_{\rm ISCO}$ and the electron temperature, $kT_{\rm e}=30\pm8$ keV changed significantly, but still comparable with the results in Table~\ref{table:spec}. Therefore, we conclude that the current model with lower solar abundance does not adequately fit the broadband spectra.

\section{Timing analysis}
\label{sec:timing}
From accurate timing analysis we can obtain a wide variety of important information on the neutron star (binary) system, including its accretion disk (during outburst), and its orbital- and spin evolution as well as on the physical processes involved in the generation of the pulsed emission. 

The first step in the timing analysis is the conversion of the on-board event arrival time expressed in MJD for the Terrestial Time system (TDT or TT) to the solar-system barycenter arrival time for the TDB time system. For this process we used a) the JPL DE405 solar system ephemeris; b) the instantaneous spacecraft location with respect to the Earth's center, and finally c) the (currently) most accurate source location of \psrtar\ as obtained by \cite{2011ATel.3606....1C}  using \cxo\ soft X-ray data (see Table \ref{table:eph}). In the event selection procedure, we ignored time intervals during which type I X-ray bursts occurred or high-background rates were encountered (see Sect.~\ref{sec:nicer}).

\begin{figure*}[t]
\centering
\includegraphics[width=18cm]{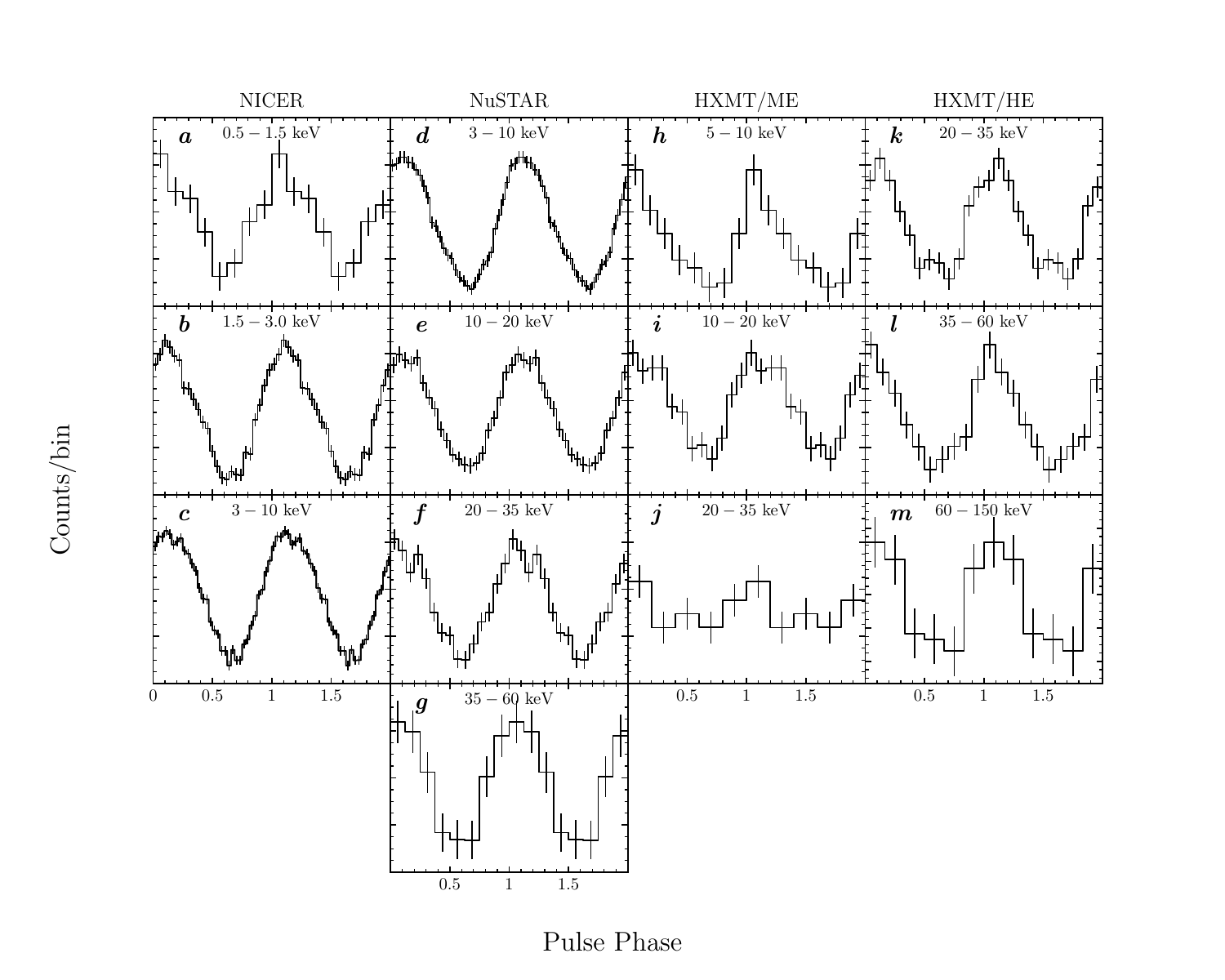}
\caption{The 0.5--150  keV broadband pulse-phase distributions of \psrtar\ observed by \nicer\ (panels $\textbf{\textit{a}}$--$\textbf{\textit{c}}$, 0.5--10 keV), \nustar\ (panels $\textbf{\textit{d}}$--$\textbf{\textit{g}}$, 3--60 keV), and \hxmt\ (panels $\textbf{\textit{h}}$--$\textbf{\textit{j}}$, 5--35 keV for ME; panels $\textbf{\textit{k}}$--$\textbf{\textit{m}}$, 20--150 keV for HE). 
Two cycles are shown to improve clarity.  The error bars represent $1\sigma$ errors. The morphology is almost unchanged with energy. All profiles reach their maximum near phase $\sim0.1$.}
\label{fig:pulse_profile}
\end{figure*}

\subsection{\nicer, \nustar, \hxmt\ and \Integ}
\label{sec:timing_general}
Because \nicer\ monitoring observations during the 2023 outburst provided the most uniform and sensitive exposure to \psrtar\ we used this set to construct an accurate timing model of the binary milli-second pulsar. The pulsed signal strength, evaluated through the bin-free $Z_{1,2}^2$-test statistics \citep{buccheri1983}, is a function of four parameters assuming a constant spin rate of the neutron star and a circular orbit (eccentricity $e\equiv0$).
We employed a 4d optimisation scheme based on a downhill {\tt SIMPLEX}
algorithm by iteratively improving the $Z^2$-statistics with respect to four parameters: the spin frequency $\nu$, the projected semi-major axis of the neutron star $a_{\rm x}\sin i$, the orbital period $ P_{\rm orb} $ and the time-of-ascending node $T_{\rm asc}$ \citep[see e.g.][for earlier (lower) dimensional versions of the method]{deFalcoa,ZLi21}\footnote{The downhill {\tt SIMPLEX} method is an optimisation algorithm to find the global minimum of a multi-parameter function. The statement in the second paragraph of Sect.~5 in \citet{deFalcoa} means to find the global minimum of $-Z_1^2$-test statistic, and so to obtain the maximum of the $Z_1^2$ distribution.}

As start values for this optimisation process (not a blind search) we used the system parameters of \psrtar\ as derived for the previous 2011 outburst \citep[see e.g.][and references therein]{falanga12,Papitto11}, except for the $T_{\rm asc}$ parameter which we adapted for current epoch using the 2011 orbital period value. The optimised values and their $1\sigma$ uncertainties, shown in Table \ref{table:eph}, are based on the first eighteen \nicer\ observations covering MJD 60054.301--60067.520 and totaling 69.425 ks of (good time/screened) exposure collected during the ON phase of the 2023 outburst.
With the accurate spin- and orbital parameters given in Table \ref{table:eph} we can calculate for any high-energy instrument used in this work the (spin) pulse-phase of each selected event taking into account the orbital motion of the pulsar, and so construct pulse-phase distributions for various energy bands within the instrument bandpass.

\begin{table}[t] 
{\small
\caption{The orbital and spin parameters of \psrtar\ as derived in this work from a 4d-optimisation scheme using \nicer\ 1.5-10 keV data. The used position \citep[see][]{2011ATel.3606....1C} in the barycentering process is shown as well. }
\centering
\begin{tabular}{lcc} 
\hline \hline 
Parameter                      & Values                           &Units    \\
\hline 
\noalign{\smallskip}  
$\alpha_{2000}$                & $17^{\hbox{\scriptsize h}} 49^{\hbox{\scriptsize m}} 55\fs35$ &         \\   
$\delta_{2000}$                & $-29\degr19\arcmin19\farcs6$   &         \\        \noalign{\smallskip}  
     
$ e $                          & $0$ (fixed)                    &         \\             
$ P_{\rm orb} $                & 13835.6176(72)                 &s        \\             
$ a_{\rm x}\sin i$             & 0.365\,181(24)                   &lt-s     \\             
$T_{\rm asc} $                 & 60053.925\,446\,1(39)              &MJD (TDB)\\             
\noalign{\smallskip}  
\hline
\noalign{\smallskip}  
\multicolumn{3}{c}{Constant Frequency model}\\
\noalign{\smallskip}  
\hline 
\noalign{\smallskip}  
Validity range                 & 60054-60067                    &MJD (TDB)\\                      
$t_0$ (Epoch)                  & 60061                          &MJD (TDB)\\
$\nu$                          & 400.990\,186\,000(25)              & Hz      \\
JPL Ephemeris                  & DE405                          &         \\           
\hline 
\end{tabular}
\label{table:eph} 
}
\end{table} 

A compilation of pulse-phase distributions (outburst integrated) for different high-energy instruments in different energy bands is shown in Fig. \ref{fig:pulse_profile}. Pulsed emission is significantly detected from $\sim$0.5~keV (\nicer) up to $\sim$150~keV (\hxmt/HE) with a 60--150~keV pulse significance of $\sim$ 3.5$\sigma$ (see Fig. \ref{fig:pulse_profile}{\it m}). The pulse-shape is approximately sinusoidal and remarkably stable as a function of energy.

Also \Integ-IBIS-ISGRI (not shown in Fig.~\ref{fig:pulse_profile}) has detected pulsed emission from \psrtar\ in a combination of data from observations taken during \Integ\ revolutions 2630, 2632, and 2634 (MJD 60053.282-60063.761) with at the highest energies a 55--90~keV pulsed emission significance of $\sim$3.5$\sigma$. {\bf{Remarkably}}, no pulsed emission (20--60 keV) was detected at the early stage of the outburst during the revolutions 2628 \citep[discovery revolution; see][]{2023ATel15996} and 2629 from MJD 60047.181--60050.708 when the source flux was rising to its (first) maximum (see Fig.~\ref{fig:outburst}, and also \citealt{papitto20}).

\begin{figure*}[t]
\centering
\includegraphics[width=7.5cm,height=7.5cm]{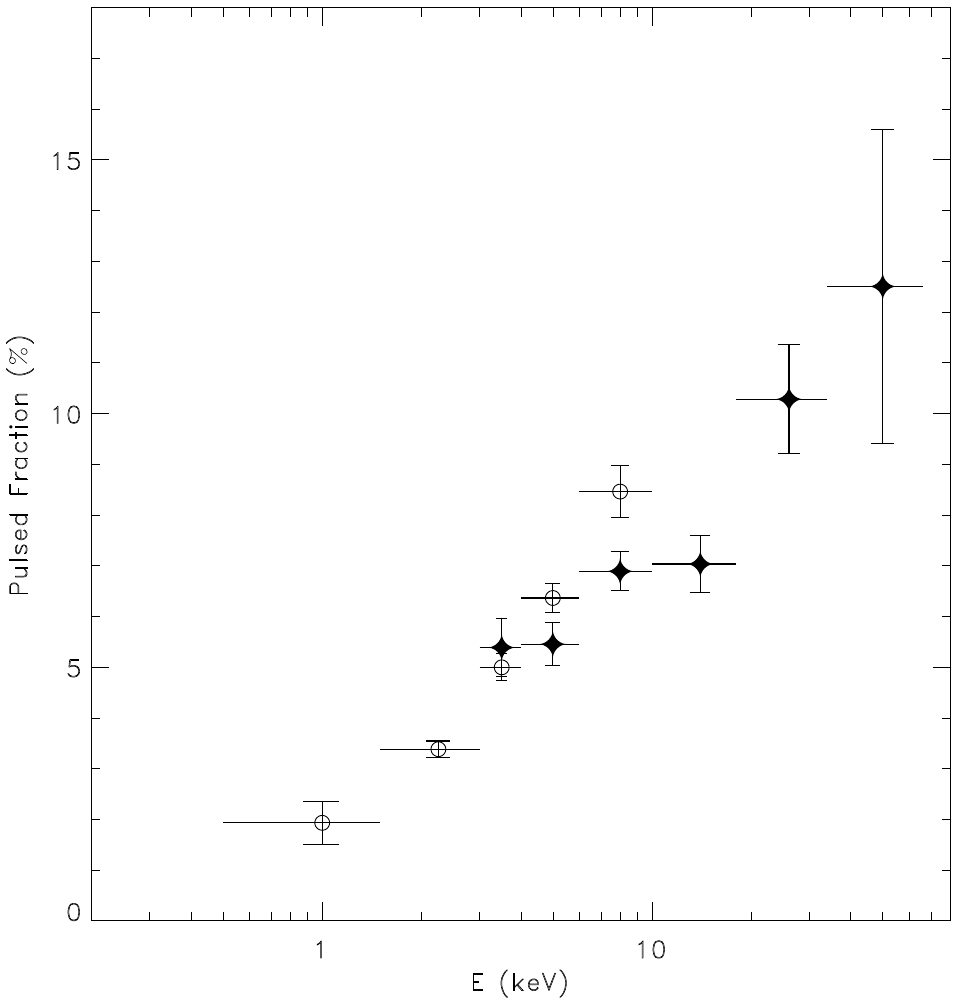}
\hspace{1cm}
\includegraphics[width=7.5cm,height=7.5cm]{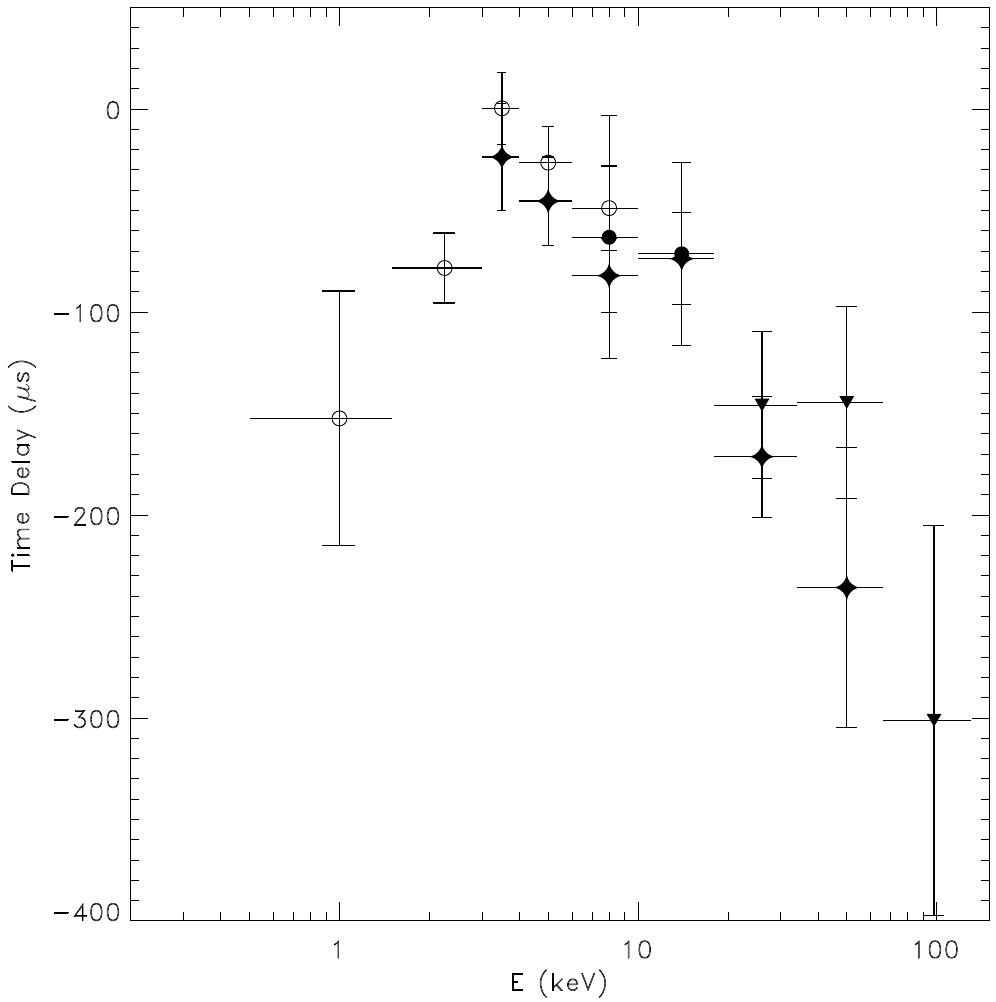}
\caption{Left: Background subtracted pulsed fraction as a function of (measured) energy using (outburst averaged) \nicer\ data (open circles) across the 0.5--10 keV band and \nustar\ data (filled diamonds) across the 3--66~keV range. The pulsed fraction clearly increases as a function of energy going from $\sim$2\% near 1~keV to $\sim$13\% at the 34--66~keV band. Right): The time delay (in $\mu$s) relative to the \nicer\ 3--4 keV band showing \nicer\ (0.5--10 keV; open circles), \nustar\ (3--66 keV; filled diamonds), \hxmt/ME (6--18 keV; filled circles) and \hxmt/HE (18--130 keV; filled triangles) measurements. Events with energies above 4 keV systematically arrive earlier with an increasing trend as a function of energy than those from the reference 3--4 keV band. Also (\nicer) events from the soft band with energies below 3 keV (not properly accessible during the 2011 outburst) do so.}
\label{fig:pulsechar}
\end{figure*}

\subsection{Variability of the pulsed emission: signal strength and pulse arrival}
\label{sec:pulse_variability}
Correcting the barycentered \nicer\ event arrival times for the orbital motion induced delays we studied the pulsed emission strength across the course of the outburst using the $Z_1^2$-test statistics as proxy for the signal-to-noise ratio \citep{buccheri1983}.  
We identified a $\sim$2~d duration episode on MJD 60060.03--60061.92 with 8.594~ks of GTI exposure exhibiting strongly suppressed pulsed emission with a $Z_1^2$ signal strength value of $4.7\sigma$. This episode coincides with the general maximum in the \nicer\ lightcurve shown in Fig.~\ref{fig:outburst}.
\nicer\ observations performed just before (MJD 60059.015--60059.975; 8.884 ks) and after (MJD 60062.225--60062.944; 4.508 ks) this episode showed pulsed-signal strength of $20.6\sigma$ and $30.7\sigma$, respectively.
Apparently, the pulsed signal was quenched during the period showing maximum accretion rate. It is interesting to note that during the \nicer\ observation (MJD 60059.015--60059.975) preceding the `quenched' episode the spin-frequency was about $3.4(5)\times 10^{-6}$ ($\sim7\sigma$) larger than the outburst averaged value shown in Table \ref{table:eph}, likely indicating a short duration spin-up period just before reaching maximum luminosity. Near the end of the outburst at MJD 60066.098--60067.520, a similar, but less pronounced ($5.6\sigma$ in 2.871~ks) episode of quenched pulsed emission is found in \nicer\ data, also coincident with a (local) maximum in the outburst lightcurve.

Next, we investigated the stability of the (binary motion corrected) pulse-arrival times by applying a time-of-arrival (ToA) analysis \citep[see e.g.][for more details]{kuiper2009}. This method assumes a stable invariant pulse shape of \psrtar\ across the outburst in the correlation process with a high-statistics template. This method applied to \hxmt\ HE 20--60 keV data yielded seven ToA measurements across MJD 60057--60071\footnote{Note that \hxmt\ HE data showed pulsed emission up to MJD 60071.389 i.e. beyond the last \nicer\ observation  ending at MJD 60067.325 during the ON phase of the outburst.} that scattered slightly around the (outburst averaged spin-frequency) model prediction. Minimizing the \hxmt\ HE ToA phase residuals yielded a spin frequency of $\nu=400.990\,185\,94(3)$Hz with a best fit root-mean-square (RMS) of 0.023 in phase units ($\equiv 57.2\ \mu$s in time domain), consistent at a $2\sigma$ level with the (outburst averaged) value given in Table \ref{table:eph}.

The ToA method applied to fifteen 1.5--10 keV \nicer\ measurements yielded a RMS of 0.092 in phase units ($\equiv 228\mu$s), which is sufficiently small to guarantee phase coherence across the outburst, but the scatter is too large to be explained by statistical fluctuations. Apparently, systematics due to accretion induced processes play an important role in the soft X-ray regime, and result
in rather large fluctuations in the pulse-arrival times.

\subsection{Pulsed fraction and time lag}
\label{sec:pulse_fraction_lag}
From the pulse-phase distributions in different (measured) energy bands (see e.g. Fig. \ref{fig:pulse_profile}), we can deduce some important intrinsic emission properties of the pulsed emission providing clues to the emission mechanism.  Focusing now on \nicer\ and \nustar, because for these instruments we can perform reliable background subtraction contrary to the \hxmt\ instruments given the weakness of the outburst, we can estimate the intrinsic (background subtracted) pulsed fraction as a function of energy band. The measured pulse-phase distributions (for a selected energy interval) can be described in terms of a truncated Fourier series,
\begin{equation}
F(\phi) = A_0 + \sum_{k=1}^2 A_k \ \cos[2\pi\ k (\phi-\phi_k)],
\label{eq10}
\end{equation}
where  $A_0$ is a constant, $A_{1}$ and $A_{2}$ are the amplitudes,  $\phi_1$ and $\phi_2$ are the phase angles in units of radians/$2\pi$, of the fundamental and the first overtone, respectively.
We can determine the global minimum for each energy band, and from this along with the total number of background subtracted source counts the intrinsic pulsed fraction. The result of this procedure is shown in the left panel of Fig. \ref{fig:pulsechar}. It is clear from this plot that the intrinsic pulsed fraction steadily increases from $\sim 2\%$ to $\sim 13\%$ going from soft X-rays at $0.5$ keV (\nicer; open circles) up to hard X-rays at $66$ keV (\nustar; filled diamonds). This pronounced trend was not visible in the 2011 outburst data, being consistent with constant \citep[cf. Fig. 7 of][]{falanga12} and having poor statistical quality in the hard X-ray regime above $\sim 20$ keV. 

In this work, we also determined the time lag as a function of energy with respect to a chosen reference energy band by applying a cross-correlation method \citep[cf. Fig. 5 of][]{falanga12}. For this purpose, we used data from \nicer, \nustar\, and also \hxmt-ME/HE, because background subtraction is not required, to obtain this quantity. The results using \nicer\ 3--4~keV band as reference interval in the cross-correlation procedure are shown in the right panel of Fig.~\ref{fig:pulsechar}.
Events having energies above 4~keV arrive earlier than the events from the 3--4~keV reference band in a decreasing way the higher the energy. The trend above $\sim$20~keV was not visible in the 2011 outburst data because of poor statistical quality. Events having energies below 3~keV, an interval that could not be studied before, also arrive earlier than the reference band. 

Thanks to the detection of broadband pulse emissions from \psrtar, the observed time lag is unique compared with other AMXPs. For most AMXPs, the time lag is constant below $\sim$2--3~keV, and monotonically increasing with energy (the soft-energy pulsed photons lag behind the hard-energy ones) and saturating at about 6--20~keV \citep[see e.g.][]{gp05,falanga11}.  The behaviors of time lag have been explained by the two-component model, i.e. the soft blackbody component from a hot spot on the NS surface, and the hard Comptonized component from the up-scattering of the seed photons from hot spot by the accretion flow \citep{Poutanen03,gp05,ip09}. The soft pulsation is dominated by the blackbody component, which vanishes above the saturated energy, i.e. higher saturated energy corresponding to a stronger contribution of the blackbody component in the total flux or higher blackbody temperature \citep{falanga11}. However, we do not observe the saturated energy till 100~keV in \psrtar. Therefore, the hard and soft time lags observed in \psrtar\ cannot be explained by this model.  The Comptonized  component may have different patterns at different energy ranges.

\begin{figure*}
\centering
\includegraphics[width=\textwidth]{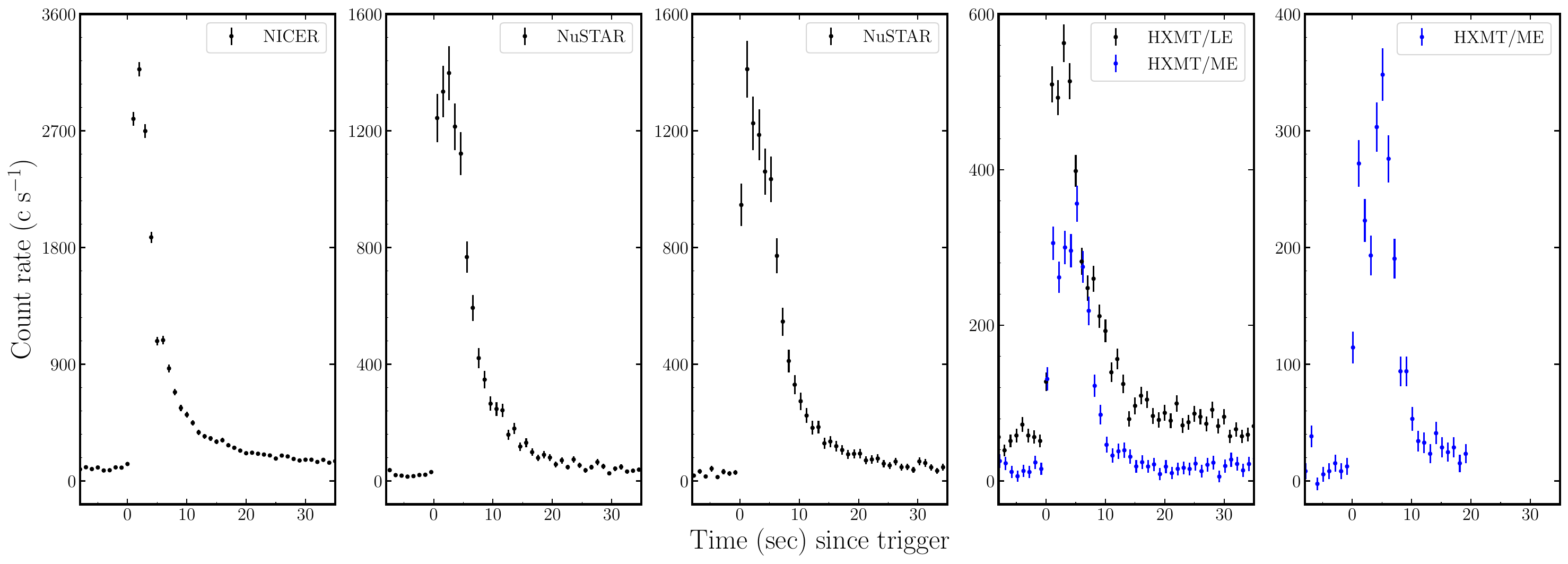}
\caption{Light curves of five X-ray bursts from \psrtar\ detected by \nicer\ (burst \#1), \nustar\ (\#2 and \#3), and \hxmt\ (\#4 and \#5). The trigger time of each burst is listed in the fourth column of Table~\ref{table:burst}.}
\label{fig:burst_lc}
\end{figure*}

\section{Type-I X-ray bursts}
\label{sec:bursts}

From 1-s binned light curves, we identified five type I X-ray bursts, one from \nicer\ (\#1), two from \nustar\ (\#2 and \#3), and two from \hxmt\ (\#4 and \#5).  The burst light curves plotted in Fig.~\ref{fig:burst_lc} all have similar profiles. For burst \#5, only ME data are available. The burst rise time is defined as the interval between the first point when the flux exceeds that of the persistent emission by 5\%, $t_{5\%}$, and the point when the count rate reaches 95\% of that in the first peak, $t_{95\%}$. For each burst, we applied a linear interpolation to the light curve to determine $t_{5\%}$ and $t_{95\%}$.  The uncertainties in the rise time are estimated from the standard deviation of the peak rate of the light curve. All bursts showed a fast rise time of $\sim$2~s. Bursts \#4 and \#5 exhibited two peaks in the  LE and ME data, and other bursts had only one peak. The flux from all bursts decreased exponentially to the pre-burst level. We searched for the burst oscillation in the frequency range of 396--406~Hz by applying $Z^2_1$ statistic \citep{buccheri1983}. The selected energy ranges are 0.5--10~keV for the \nicer\ burst, 3--20~keV for two \nustar\ bursts, and 5--20~keV for two \hxmt\ ME bursts. No significant oscillation signal has been detected. 

The time-resolved spectra were extracted with an exposure time of 1--8~s, which ensured that the spectra had sufficient counts to perform spectral fitting and avoided missing the PRE burst features. We fit the burst spectra with an absorbed blackbody model, while the pre-burst spectra were regarded as background and assumed unchanged during bursts. We fixed the hydrogen column density, $N_{\rm H}$, at the values obtained in Sect.~\ref{sec:nicer_spec}. The free parameters are the temperature, $kT_{\rm BB}$, and the normalization, $K=R^2_{\rm ph}/D^2_{\rm 10~kpc}$, of the blackbody component. Most burst spectra were well fitted yielding $\chi^2_{\nu}$ of $\sim1.0$.   We identified three PRE bursts from the time-resolved spectroscopy, i.e. the first burst from  \nustar\ and two bursts from \hxmt. Assuming a distance of 8~kpc the radii of the photosphere expanded to $13.5\pm0.3$, $11.3\pm0.7$ and $9.3_{-1.4}^{+2.2}$~km, with corresponding temperatures of $1.9\pm0.1$, $2.0\pm0.1$ and $2.2\pm0.2$~keV, for bursts \#2, \#4 and \#5, respectively. The peak fluxes of all bursts are around $(4.0\pm0.5)\times10^{-8}~{\rm erg~s^{-1}~cm^{-2}}$, which are lower than the brightest PRE burst reported by \citet{falanga12}. 

We fitted the exponential decay of the bursts with the function $F(t)=F_{0}\exp(-t/\tau)$, where $\tau$ is the decay time, and found that all bursts had decay times close to $5$ s. Next, the burst fluence was derived using the relation $f_{\rm b}=F_{\rm peak}\tau$. The bursts' short rise and decay times of our 2023 sample are consistent with those from the sample of the 2011 outburst, which were powered by unstable burning of hydrogen-deficient material. The properties of the 2023 outburst bursts, the start time, the rise time, the peak flux, the persistent flux, the decay time $\tau$, and the burst fluence, are listed in Table~\ref{table:burst}. 

We calculated the burst recurrence times and obtained 50.1, 9.1, 37.7 and 27.5 hrs starting from the burst onset time. The pair of two \nustar\ bursts had the shortest recurrence time. We tentatively considered this as the true recurrence time, i.e. no X-ray burst(s) were missed between these two bursts, which is shorter than the recurrence time of $\sim$16--18~hr found during the 2011 outburst \citep{falanga12}. The other recurrence times were 5.5, 4 and 3 times longer, indicating that some bursts could have been missed due to the data gaps. 


The burst persistent fluxes were obtained from interpolating the flux reported in Sect.~\ref{sec:nicer_spec}. All uncertainties were assigned as their typical values. The X-ray bursts were triggered during the outburst with the persistent flux in the range of $(0.97-1.34)\times10^{-9}~{\rm erg~s^{-1}~cm^{-2}}$. Assuming a distance of 8 kpc, the persistent flux corresponds to $(0.74-1.03)\times10^{37}~{\rm erg~s^{-1}}$, or $(2.0-2.7)\%L_{\rm Edd}$ by using the empirical value $L_{\rm Edd}=3.8\times10^{38}~{\rm erg~s^{-1}}$ \citep{kuulkers03}. The local accretion rate per unit area is calculated from the relation $\dot{m}=L_\mathrm{{per}}(1+z)(4\pi R_{\rm NS}^{2}(GM_{\rm NS}/R_{\rm NS}))^{-1}$, i.e. $\dot{m}\sim(4.15-5.72)\times10^4~{\rm g~cm^{-2}~s^{-1}}$, assuming the mass, $M_{\rm NS}=1.4M_\odot$, and radius, $R_{\rm NS}=10$ km, of NS, and thus the gravitational redshift $1+z=1.31$. From the observed burst fluences, the total burst released energies are $E_{\rm b}=4\pi d^2f_{\rm b}=(1.1-2.0)\times10^{39}$ erg. We estimated the burst ignition depth via 
$y_{\rm ign}=4\pi E_\mathrm{{b}}d^{2}(1+z)(4\pi R_{\rm NS}^{2}Q\mathrm{_{nuc}})^{-1}$, 
where the nuclear energy generated for pure helium is $Q_\mathrm{{nuc}} \approx 1.31 \mathrm{~MeV~nucleon^{-1}}$ for $X=0$  \citep{Goodwin19}. We obtained $y_{\rm ign}=(0.96-1.66)\times10^8~{\rm g~cm^{-2}}$.
Once the ignition depth is known, we calculated the recurrence time between bursts by using the equation $\Delta t_{\mathrm{rec}}=(y_{\mathrm{ign}} / \Dot{m})(1+z)$. The predicted recurrence time is $(6.4-11.0)$ hr, consistent with the observed values.



\begin{figure*}
\centering
\includegraphics[width=\textwidth]{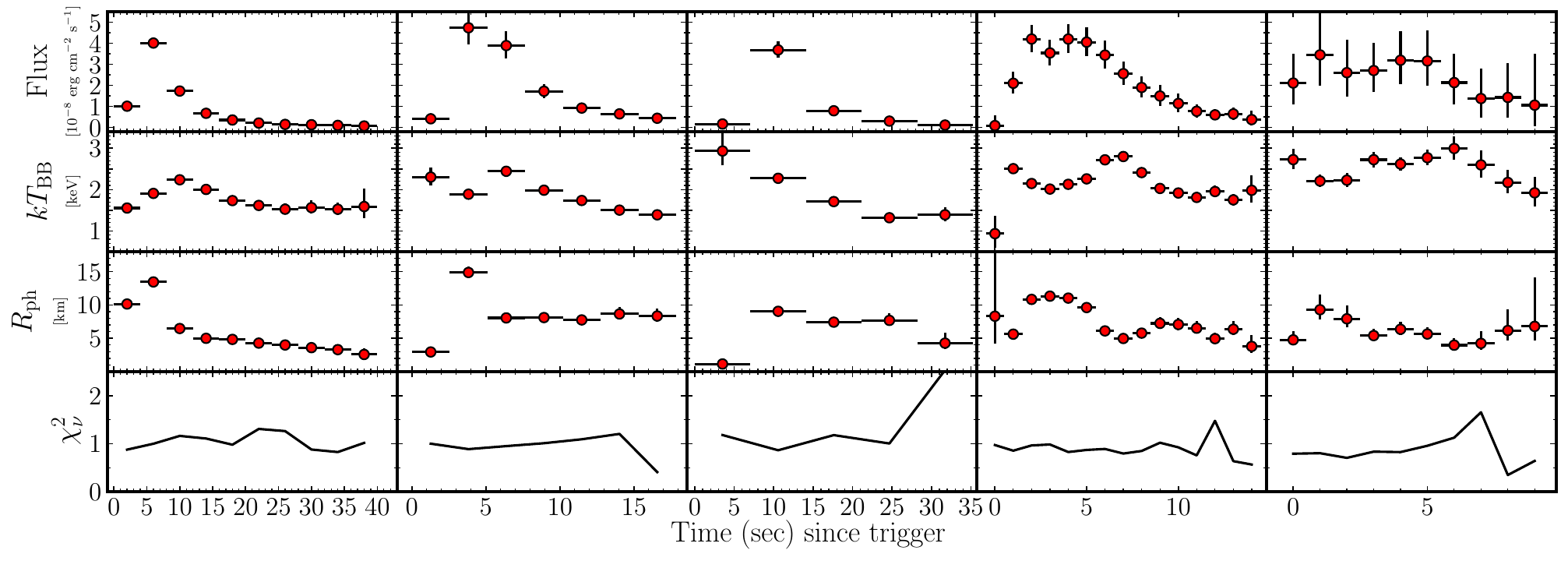}
\caption{Evolution of the spectral parameters of five X-ray bursts. From left to right panels, the bursts are shown in the same order as Fig.~\ref{fig:burst_lc}.}
\label{fig:burst}
\end{figure*}

\begin{table*}
\centering
\small
\caption{The \psrtar\ burst properties for the 2023 outburst sample. 
}
\begin{tabular}{lcccccccc} 
\hline \hline
Burst ObsID & Instrument & PRE  & Start Time  &   Rise time    & Peak Flux & Persistent flux &$\tau$ & $f_{\rm b}$ \\
& & & MJD  & s   & $10^{-8}~{\rm erg~s^{-1}~cm^{-2}}$&$10^{-9}~{\rm erg~s^{-1}~cm^{-2}}$ & s &   $10^{-7}~{\rm erg~cm^{-2}}$\\ 
\hline 
\noalign{\smallskip}  
6560010101 &\nicer & N &  60055.71810  &  $1.8\pm0.1$    & $4.01\pm0.11$ & $1.28\pm0.05$ &  $4.78\pm0.19$  & $1.92\pm0.94$ \\
90901317002 & \nustar &Y &60057.80410  &  $2.0\pm0.4$     & $4.74\pm0.78$ & $1.34\pm0.01$ &  $5.41 \pm 1.03$   &$2.56 \pm 0.65$ \\
90901317002 & \nustar &N &60058.18364  &  $1.8\pm0.1$    & $3.68\pm0.38$ & $1.34\pm0.01$  &  $4.79 \pm 0.43$   &$1.76 \pm 0.24$ \\
P050409300302 & \hxmt\ &Y &60059.75634  &  $2.0\pm0.3$    & $4.20\pm0.62$ & $0.97\pm0.05$  &  $4.58 \pm 0.39$ &
$1.92 \pm 0.33$ \\
P050409300402 & \hxmt\ &Y  &60060.90046  &  $2.0\pm0.4$   & $ 3.45\pm1.46 $&$1.28\pm0.05$ &  $4.28 \pm 0.97$  & 
$1.48 \pm 0.71$ \\

\noalign{\smallskip}  
\hline  
\end{tabular}  
\label{table:burst}
\end{table*}



\section{Discussion and summary}
\label{sec:disc}

In this work we have studied the X-ray pulsations, time-resolved and broadband spectra, and X-ray burst properties of the AMXP \psrtar\ during its 2023 outburst. 
Five type I X-ray bursts have been detected, including three PRE bursts. These X-ray bursts exhibited similar profiles, with the rise time of $\sim$2~s, exponential decay time of $\sim$5~s, and the peak flux of around $4\times10^{-8}$~erg~s$^{-1}$~cm$^{-2}$.
The peak fluxes of the PRE bursts are lower than that from the brightest burst observed by RXTE \citep{falanga12}. We measure a recurrence time of 9.1~hr, shorter than the values, $\sim$16--18~hr, observed during its 2011 outburst at a similar persistent flux.  These properties can be well explained by the unstable burning of hydrogen-deficient material on the NS surface in \psrtar.

\subsection{Broadband spectra}

We analyzed the spectra observed from \nicer, \nustar, \hxmt and \Integ\ IBIS-ISGRI, covering the 1--150~keV energy range. The \nicer\ spectra in 1--10~keV  can be well fitted with an absorbed \texttt{nthcomp} plus \texttt{Gaussian} model with $N_{\rm H}\sim(2.4-2.9)\times10^{22}~{\rm cm^{-2}}$, a power-law photon index $\Gamma$ of $ \sim$1.6--1.8, an electron temperature $kT_{\rm e}$ of $\sim$2.9--9.4~keV, and a seed photon temperature $kT_{\rm BB}$ of $\sim$0.25--0.46~keV.  The bolometric flux reached its peak around $1.35\times10^{-9}$~erg~s$^{-1}$~cm$^{-2}$ in $\sim$2--3~d, similar to its 2011 outburst and also to other AMXPs. However, the source showed fluctuations in the next 10~d in both the soft and hard X-ray bands, which is peculiar considering that typically a smooth decay is observed for most of the AMXP outbursts \citep{falanga05,Kuiper20,Li23}. 
During the decay stage a reflare appeared.

\citet{Marino19} calculated the long-term average luminosity of \psrtar, assuming an orbital evolution driven by conservative mass transfer, by considering two possible mechanisms: gravitational radiation (GR) and magnetic braking (MB). They determined that the average luminosity was $1.8 \times 10^{35}~{\rm erg~s^{-1}}$ for the GR-only case and $3.7 \times 10^{35}~{\rm erg~s^{-1}}$ for the GR+MB case. However, the observed average X-ray luminosity before the 2023 outburst was significantly lower, at $0.61 \times 10^{33}~{\rm erg~s^{-1}}$, which is far below the theoretically predicted values. As a result, they suggested that non-conservative mass transfer likely occurred during the 2011 outburst of \psrtar. The 2023 outburst profile was close to that of 2011. Adding an extra outburst, the observed average luminosity over the last 20 years has doubled, but it remains below the predictions. Therefore, the 2023 outburst of IGR J17498--2921 likely involved non-conservative mass transfer as well.



The joint quasi-simultaneous \nicer, \nustar and \Integ\ IBIS-ISGRI spectra in the energy range of 1--150~keV are well described by a self-consistent reflection model, \texttt{relxillCp}, with a Gaussian line, modified by interstellar absorption. The unabsorbed bolometric flux $F_{\rm bol}$ of the broadband spectrum was $(1.45\pm0.01)\times 10^{-9}$~erg~s$^{-1}$~cm$^{-2}$, corresponding to $\sim3\%L_{\rm Edd}$ assuming a source distance of 8~kpc. We can characterize the accretion flow properties by a photon index of  $1.78\pm0.01$ and an electron temperature of $kT_{\rm e}=39\pm4$~keV. The inclination angle of the binary system is $34\pm3$~deg. We found a high reflection fraction of $f_{\rm refl} = 5.2_{-0.7}^ {+1.0}$, which means most of the photons interacted with the surrounding disk first rather than emitted directly to the observer. It is self-consistent with our first measurement of the inner disk radius $R_{\rm in}= 1.7_{-0.3}^{+0.4}\ R_{\rm ISCO}$ for \psrtar. The broadband spectral fitting shows the properties of the accretion disk as a strong ionization, $\log(\xi/{\rm erg~cm~s^{-1}})=4.46_{-0.07}^{+0.14}$, over-solar abundance, $A_{\rm Fe}=7.7_{-0.9}^{+1.7}$, and a high density, $\log (n_{\rm e}/{\rm cm^{-3}})=19.5_{-0.5}^ {+0.3}$. There are several X-ray binaries showing over-solar abundance in the spectra, which can be alternatively explained by a reflection model with a higher disk density \citep[see e.g.,][]{Ludlam17,Ludlam18,Ludlam19,Tomsick18,Connors21,Liu23}. In the case of \psrtar,  we note that  $A_{\rm Fe}$ and $\log \xi$ present significant negative covariances with $\log n_{\rm e}$ in Fig.~\ref{fig:mcmc}. It is not surprising because the ionization parameter $\xi$ is related to the accretion density  $n_{\rm e}$ via $\xi = 4\pi F_{\rm x}/n_{\rm e}$, where $F_{\rm x}$ is the total illuminating flux.  For the persistent luminosity around a few percent of the Eddington limit, the disk density can reach up to $\sim10^{22}$~cm$^{-3}$ \citep{Jiang19}. The upper limit of $\log n_{\rm e}$ in the current \texttt{relxillCp} model is 20. Therefore,  a larger disk density with smaller ionization and solar abundance may also be possible. We also found that the important parameters, $R_{\rm in}$ and $i$, weakly depend on the disk density. Thus, a higher $n_{\rm e}$ does not affect our measurements of these two parameters. From the pulsar mass function $f(M_{\rm NS}, M_{\rm  C}, i)=M_{\rm C}^3\sin^3i/(M_{\rm NS}+ M_{\rm  C})^2\approx 2\times10^{-3} M_\odot$, we calculate the companion star mass $M_{\rm C}=$(0.32--0.41)$M_\odot$ for $M_{\rm NS}\sim$1.4--2$M_\odot$ if the accretion disk is aligned with the binary orbit.

\subsection{Orbital period refinement}

Comparing the orbital solutions obtained for the 2011 and 2023 outbursts, we can improve the orbital period $P_{\rm orb}$. 
The integer number of orbital cycles between $T_{\rm asc}^{2023}=60053.9254461$ MJD and $T_{\rm asc}^{2011}=55786.18099710$ MJD from current work and \citet{falanga12}, respectively, amounts $N_{\rm cyc}=$ int $\left((T_{\rm asc}^{2023}-T_{\rm asc}^{2011})/P_{\rm orb}\right)=26651$. 
In turn, assuming that the orbital period between two outbursts is unchanged, we can refine the orbital period to $P_{\rm orb}=(T_{\rm asc}^{2023}-T_{\rm asc}^{2011})/N_{\rm cyc} = 13835.620442(13)$ s, where the error is mainly from the uncertainty in $T_{\rm asc}^{2023}$ given in Table~\ref{table:eph}. 

Constraints on the upper and lower limits of $\dot P_{\rm orb}$ are also possible,  even though a direct measurement is impossible with only two $T_{\rm asc}$ measurements.  Using the equation from \citet{Burderi09} or \citet{Riggio11}, $\dot P_{\rm orb}$ is expressed as
\begin{equation}
    \dot P_{\rm orb}=\frac{2}{N_{\rm cyc}P_{\rm orb}}\left(\frac{\Delta T_{\rm asc}(N_{\rm cyc})}{N_{\rm cyc}}-\Delta P_{\rm orb}\right),
\end{equation}
where $\Delta T_{\rm asc}(N_{\rm cyc})=T_{\rm asc, ~2023}-(T_{\rm asc, ~2011}+N_{\rm cyc}\times P_{\rm orb})$ is the difference between the predicted and measured time of the ascending node in 2023. If we adopt the upper and lower limits of the orbital period $P_{\rm orb}=13835.619(1)$~s and $\Delta P_{\rm orb}=0.001$~s measured by  \citet{Papitto11}, we obtain  $\dot P_{\rm orb}$ of $(-3,+8)\times10^{-12}$~s~s$^{-1}$ at $1\sigma$ confidence level.

\subsection{The stellar magnetic field and the long-term spin evolution}

If we assume that the inner accretion disk is truncated at the magnetospheric (Alfvén) radius, the magnetic dipole moment can be expressed as \citep{Psaltis99,ip09},
\begin{equation}\label{equ:mag1}
\begin{split}
\mu_{26}=0.128\times k^{-4/7}_{\rm A} 
\left(\frac{M_{\rm NS}}{1.4M_\odot}\right)^{1/4}\left(\frac{R_{\rm in}}{10~ {\rm km}}\right)^{7/4}\times \\ 
\left(\frac{f_{\rm ang}}{\eta}\frac{F_{\rm per}}{10^{-9}~{\rm erg~s^{-1}~cm^{-2}}}\right)^{1/2}\frac{D}{8~{\rm kpc}},
\end{split}
\end{equation}
where $\mu_{26}=\mu/10^{26}~{\rm G~cm^3}$, $\eta$ is the accretion efficiency,  the conversion factor $k_{\rm A}$ and the angular anisotropy factor $f_{\rm ang}$ are set to unity, the distance $D$ to the source is 8 kpc, the bolometric flux and the inner disk radius from Sect.~\ref{sec:nicer_nustar} are used. If we adopt $\eta\sim0.1-0.2$, the NS mass $M_{\rm NS}\sim1.4-2M_\odot$ and radius $R_{\rm NS}\sim{\rm}\sim10-15$ km, the distance uncertainty of 0.8 kpc, and the inner disk radius $R_{\rm in}=18.6_{-3.3}^{+4.5}$ km, the magnetic dipole moment is $\mu_{26}\sim0.6-2.4$. This estimation converts to the NS magnetic field of $(0.2-2.4)\times10^8$ G. 




The timing solution reported by \citet{Papitto11} is $\nu=400.99018734(1)$ Hz, $\dot{\nu}=-(6.3\pm1.9)\times10^{-14}~{\rm Hz~s^{-1}}$ and $P_{\rm orb}=13835.619(1)$ s at the reference epoch $T_0= 55786.124$ MJD during the 2011 outburst.  Considering the 2011 outburst lasting between MJD 55786.1--55826.4, we obtain a spin frequency of 400.99018712(7) at the end of the 2011 outburst. If we assume that the pulsar was spinning down at a constant rate during the quiescent state, which lasted from the end of the 2011 outburst to the beginning of the 2023 outburst, then we can compute an averaged long-term spin-down rate of $\dot\nu=-3.1(2)\times10^{-15}~{\rm Hz~s^{-1}}$, where the error is calculated by propagating the uncertainties in spin frequencies. If the average spin-down rate during the quiescent state is caused by (rotating) magnetic dipole emission, the magnetic dipole moment is \citep{Spitkovsky06}, 
\begin{equation}\label{equ:mag2}
\mu_{26}=8.27\times (1+\sin^2\theta)^{1/2}I_{45}^{1/2}\nu_2^{-3/2}(-\dot\nu_{-15})^{1/2},
\end{equation}
where $\theta$ is the angle between the rotation and magnetic axes, $I_{45}$ is the NS moment of inertia in units of $10^{45}~{\rm g~cm^2}$, the spin frequency, $\nu_2=\nu/100$ Hz, and the spin frequency derivative $\dot\nu_{-15}=\dot\nu/10^{-15}~{\rm Hz~s^{-1}}$. If we adopt the same range of NS radius as Eq.~\eqref{equ:mag1} and $\sin^2\theta\sim0-1$, we have the magnetic dipole moment $\mu_{26}\sim1.6-2.2$, corresponding to a magnetic field strength of $(0.9-4.4)\times10^8$ G. The deduced magnetic field is consistent with the value calculated from Eq.~\eqref{equ:mag1}. If we combine the results from Eqs.~\eqref{equ:mag1} and \eqref{equ:mag2} then we obtain a constrain on the magnetic field of $(0.9-2.4)\times10^8$ G, which is compatible with those of other AMXPs \citep[see e.g.,][]{Hartman2008,Patruno09c,Patruno2010,Hartman2009,Hartman11,Papitto11,Li23}, and the samples provided in \citet{Mukherjee15}.

\begin{acknowledgements}
We appreciate the valuable comments from the referee, which has contributed to improving our manuscript. This work was supported by the Major Science and Technology Program of Xinjiang Uygur Autonomous Region (No. 2022A03013-3),  the National Natural Science Foundation of China (12103042, 12273030, 12173103, U1938107), and Minobrnauki grant 075-15-2024-647 (JP).
This research has made use of data obtained from the High Energy Astrophysics Science Archive Research Center (HEASARC), provided by NASA's Goddard Space Flight Center, and also from the HXMT mission, a project funded by the China National Space Administration (CNSA) and the Chinese Academy of Sciences (CAS). 
\end{acknowledgements}

\bibliography{ms_IGRJ17498.bbl}{}
\bibliographystyle{aa}

\end{document}